\documentclass{article} 
\PassOptionsToPackage{table}{xcolor}
\usepackage{iclr2027_conference,times}

\usepackage{amsmath,amsfonts,bm}

\def\eqref#1{equation~\ref{#1}}

\def\1{\bm{1}}

\DeclareMathAlphabet{\mathsfit}{\encodingdefault}{\sfdefault}{m}{sl}
\SetMathAlphabet{\mathsfit}{bold}{\encodingdefault}{\sfdefault}{bx}{n}

\usepackage{hyperref}
\hypersetup{colorlinks=true}
\usepackage{url}
\usepackage{graphicx}
\usepackage{subcaption}
\usepackage{textcomp}
\usepackage{xcolor}
\usepackage{booktabs}    
\usepackage{tikz}
\usepackage{multirow}    
\usepackage{amsmath,amssymb,amsfonts}
\usepackage{booktabs,tabularx,array,amssymb,CJKutf8}
\usepackage{booktabs}
\usepackage{multirow}
\usepackage{graphicx}
\usepackage{xcolor}
\usepackage{natbib}
\usepackage{wrapfig}
\definecolor{questionblue}{RGB}{45,95,145}

\definecolor{QwenRow}{HTML}{EAF1FA}
\definecolor{PrivateRow}{HTML}{F5F5F5}
\definecolor{EmiliaRow}{HTML}{EAF4F1}

\title{ImageTTS: Repurposing Pretrained Image Diffusion Transformers for Speech Synthesis}

\iclrfinalcopy

\author{%
\begin{minipage}[t]{0.96\textwidth}
\centering
\textbf{Hanchen Pei\textsuperscript{1}\quad Shujie Liu\textsuperscript{2}\quad
Yayue Deng\textsuperscript{3}\quad Fangyun Wei\textsuperscript{2}}\\[2pt]
\textbf{Yanqing Liu\textsuperscript{2}\quad Jianwei Yu\textsuperscript{2}\quad
Gongping Huang\textsuperscript{1}}\\[6pt]
{\small\normalfont
\textsuperscript{1}Electronic Information School, Wuhan University, Wuhan, China\\
\textsuperscript{2}Microsoft Corporation, Redmond, US\\
\textsuperscript{3}The Chinese University of Hong Kong, Hong Kong, China\\[4pt]
\scriptsize
\texttt{phc\_whu@whu.edu.cn}\quad \texttt{shujliu@microsoft.com}\quad
\texttt{yayue.apply@gmail.com}\\
\texttt{Fangyun.Wei@microsoft.com}\quad \texttt{yanqliu@microsoft.com}\\
\texttt{tomasyu@foxmail.com}\quad \texttt{gongpinghuang@whu.edu.cn}
}
\end{minipage}%
}

\begin{document}

\maketitle
\fancyhead{}
\renewcommand{\headrulewidth}{0pt}

\begin{abstract}

Pretrained image-generation models have been successfully repurposed for various downstream visual tasks. 
Whether their learned representations and generative priors can transfer beyond vision to other modalities remains largely unresolved, particularly for speech synthesis, which requires linguistic fidelity and preservation of speaker identity from a short reference.
We present ImageTTS, which repurposes FLUX.2 for speech synthesis by encoding pseudo-RGB log-mel spectrograms with its frozen image VAE and adapting only its pretrained Diffusion Transformer (DiT) using conditional latent flow matching.
This formulation disentangles two forms of vision-to-speech transfer: representation transfer through the image VAE and generative-prior transfer through DiT initialization.
The frozen image VAE largely preserves intelligibility and speaker identity with 0.99 STOI and 0.95 speaker similarity, while image-pretrained DiT initialization accelerates convergence and improves linguistic accuracy, reducing WER from 4.2\% to 2.7\% at 100k matched updates relative to random initialization.
The same framework supports both zero-shot and instruction-guided TTS and, when scaled to 100K hours of speech data, achieves 1.12\% WER on Seed-TTS test-en and a Style-ACC of 0.66 on CapSpeech, outperforming speech-native baselines trained on comparable data across both tasks.
These results suggest that pretrained image models can transfer beyond architectural reuse, supplying useful latent representations and generative priors for intelligible and controllable speech synthesis.


\end{abstract}

\section{Introduction}



Pretrained generative models are increasingly repurposed beyond the tasks and data modalities on which they were originally trained. For image generation, diffusion and flow-based models have shown that learned representations and denoising priors can support a range of downstream visual problems, including image editing, geometric prediction, and video generation~\citep{blattmann2023stable,ke2024repurposing,fu2024geowizard}. However, nearly all successful adaptations remain within the visual domain, leaving unresolved whether a text-to-image model can provide not only a reusable architecture but also the \emph{latent representation} and \emph{pretrained generative prior} for speech synthesis.

Speech synthesis offers a demanding test of this cross-modal transfer. Unlike many generation tasks in which text specifies high-level semantics, speech synthesis requires both faithful realization of the linguistic content and preservation of speaker identity from a reference utterance.
Recent speech generation systems have adopted architectural and algorithmic innovations originating from image generation. For example, diffusion-based speech models apply iterative denoising to waveform and mel-spectrogram generation~\citep{kong2021diffwave,popov2021grad}, while F5-TTS and MaskGCT further adopted Diffusion Transformers (DiTs) and MaskGIT-style architectures, respectively~\citep{chen2025f5,peebles2023scalable,wang2025maskgct,chang2022maskgit}. Riffusion~\citep{Forsgren2022riffusion} provides an early exception by fine-tuning Stable Diffusion~\citep{rombach2022high} on spectrogram images for text-conditioned music generation, which offers initial evidence that representations and generative priors acquired from images can be useful for spectrogram modeling.

Despite this early evidence, systematic transfer of pretrained image knowledge to speech remains underexplored. We therefore study vision-to-speech transfer through three related but separable evaluations, focusing on representation transfer, initialization transfer, and downstream scaling:
\textbf{\textcolor{questionblue}{\textit{Q1:}}} Can a pretrained image latent space faithfully represent speech? \textbf{\textcolor{questionblue}{\textit{Q2:}}} Do image-generation priors improve speech synthesis relative to random initialization? and \textbf{\textcolor{questionblue}{\textit{Q3:}}} How competitive can a vision-to-speech transfer paradigm become at scale?

\begin{figure}[t]
    \centering
    \includegraphics[
        width=\linewidth,
        trim=0 0.2cm 0 0.2cm,
        clip
    ]{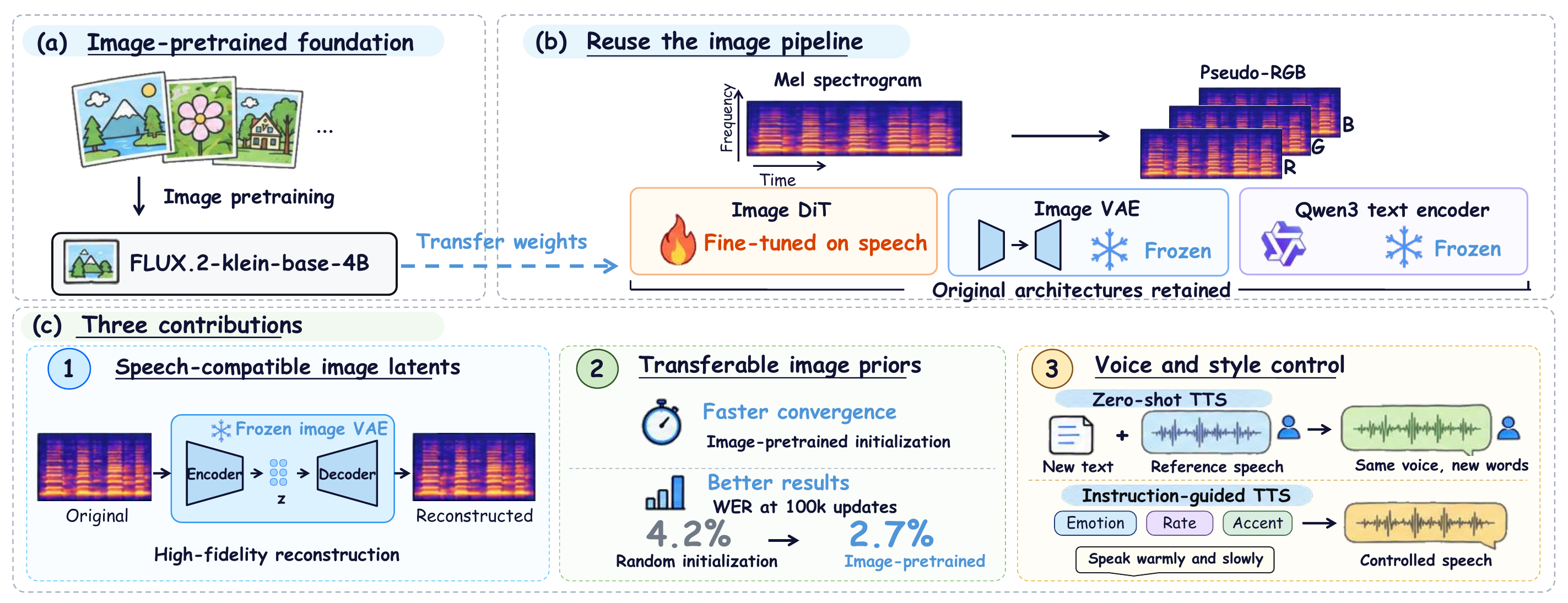}
    \caption{ Overview of ImageTTS. (a) FLUX.2 is pretrained on images and provides the DiT initialization for speech generation. (b) Log-mel spectrograms are converted into pseudo-RGB inputs; the image VAE and Qwen3 text encoder remain frozen, while the DiT is fine-tuned for speech. (c) Three findings: the image VAE supports high-fidelity speech reconstruction, image-pretrained initialization accelerates convergence and improves WER over random initialization, and the framework supports zero-shot TTS and instruction-guided TTS.}
    \label{fig:placeholder}
    \vspace{-10pt}
\end{figure}

To investigate these questions, we present ImageTTS, a framework that repurposes FLUX.2-klein-base-4B~(hereafter FLUX.2) for speech generation, as shown in Figure~\ref{fig:placeholder}. ImageTTS converts log-mel spectrograms into pseudo-RGB images, encodes them with the frozen FLUX.2 image VAE, and fine-tunes only the pretrained DiT backbone using conditional latent flow matching. Within this formulation, zero-shot TTS is cast as reference-conditioned latent infilling, whereas instruction-guided TTS is formulated as full latent generation conditioned on a transcript and a description. 
Our contributions are threefold:
\begin{itemize}
\item \textbf{A direct vision-to-speech bridge:}
We introduce ImageTTS, a controlled framework for studying transfer from a pretrained text-to-image model to speech synthesis, separating image-latent representation transfer from image-pretrained generative-prior transfer.

\item \textbf{Evidence of transfer:}
We provide evidence for transfer at both levels. The frozen image VAE preserves high speech intelligibility and speaker identity information, and image-pretrained DiT initialization improves convergence and linguistic accuracy over a randomly initialized model under matched training conditions.

\item \textbf{Scaling to zero-shot and instruction-guided TTS:}
With approximately 100K hours of speech pretraining, ImageTTS-base achieves
competitive zero-shot TTS performance. Further trained on CapSpeech, ImageTTS-instruct enables natural-language control of speaker attributes and speaking styles, outperforming comparable baselines.

\end{itemize}

\section{Related work}

\subsection{Diffusion and flow models for zero-shot TTS}
NaturalSpeech 2 was among the earliest large-scale non-autoregressive systems for zero-shot TTS, generating speech latents conditioned on explicit phoneme alignments and acoustic priors~\citep{shen2024naturalspeech2}. DiTTo-TTS subsequently introduces a scalable Diffusion Transformer (DiT), supplying variable-length text tokens through cross-attention~\citep{lee2025ditto}. A parallel line of work formulates zero-shot TTS as flow-matching-based mel-spectrogram infilling. While Voicebox establishes this formulation using frame-aligned phoneme conditions~\citep{le2023voicebox}, E2-TTS removes the external phoneme aligner by padding text tokens with fillers to match speech length~\citep{eskimez2024e2}. F5-TTS retains this equal-length conditioning scheme while replacing the U-Net-style Transformer with a DiT backbone~\citep{chen2025f5}. M3-TTS further removes filler-based pseudo-alignment by applying shared self-attention to concatenated text and speech tokens in its Joint-DiT, after which it discards the text branch and processes only the speech tokens with its Single-DiT~\citep{wang2025m3}. ImageTTS similarly avoids token-level alignment, but differs in how multimodal information is aggregated. Instead of early shared fusion followed by speech-only refinement, ImageTTS retains modality-specific representations in early double-stream blocks, followed by fully unified multimodal modeling in subsequent single-stream blocks.

\subsection{Representation spaces for speech synthesis}
Speech generation systems predominantly rely on discrete or continuous acoustic targets. Discrete codec tokens allow speech to be modeled using autoregressive or masked prediction across multiple or factorized codebooks~\citep{chen2025neural,wang2025maskgct,ju2024naturalspeech,li2024singlecodec,wang2025spark,du2024cosyvoice}. Conversely, waveform-space methods bypass intermediate tokens entirely by directly generating raw audio patches~\citep{gao2023e3tts,chen2026wavtts}, though at the expense of high temporal redundancy.
To retain continuous acoustic details without discrete quantization, models typically operate on mel-spectrograms or continuous latent tokens. Mel-spectrograms serve as fixed, continuous 2D acoustic features and are widely modeled via diffusion, flow matching, or continuous autoregressive prediction~\citep{popov2021grad,mehta2024matcha,zhu2025zipvoice,meng2025autoregressive}. To achieve higher temporal compression, other continuous methods train dedicated speech-specific autoencoders to extract continuous latent representations~\citep{jiang2025megatts3,jia2025ditar,luo2026stable}.
ImageTTS investigates an alternative continuous latent space. Instead of training a speech-specific encoder, it encodes pseudo-RGB mel-spectrograms using an off-the-shelf, frozen image VAE. This allows us to examine whether a visual latent space can effectively preserve speech information and support conditional synthesis.

\subsection{Instruction-guided TTS}

Natural-language instruction-guided TTS enables flexible control over not only what to say but also how to say it,
including speaking style, emotion, and speaker attributes. Early approaches use dedicated prompt encoders to derive style representations from natural-language descriptions and condition speech synthesis alongside the input transcript~\citep{guo2023prompttts,yang2024instructtts}. More recent systems instead place the transcript and description in a unified language interface, allowing a single instruction sequence to condition codec-token generation and support finer-grained control~\citep{zhou2024voxinstruct,chen2026flexivoice}. These developments are supported by speech datasets paired with natural-language descriptions~\citep{guo2023prompttts,ji2025controlspeech,wang2025capspeech}. ImageTTS reuses the pretrained image text encoder for both the transcript and the style description, which provides a test of whether an image-generation pipeline can understand natural-language vocal descriptions and ground them in controllable speech generation.

\section{Method}
In this section, we introduce ImageTTS. We first describe how mel-spectrograms are converted into pseudo-RGB inputs and represented in the latent space of the frozen image VAE. We then formulate speech generation as conditional flow matching in this image latent space. Finally, we instantiate this formulation for zero-shot TTS and instruction-guided TTS.

\begin{figure}
    \centering
    \includegraphics[width=1\linewidth]{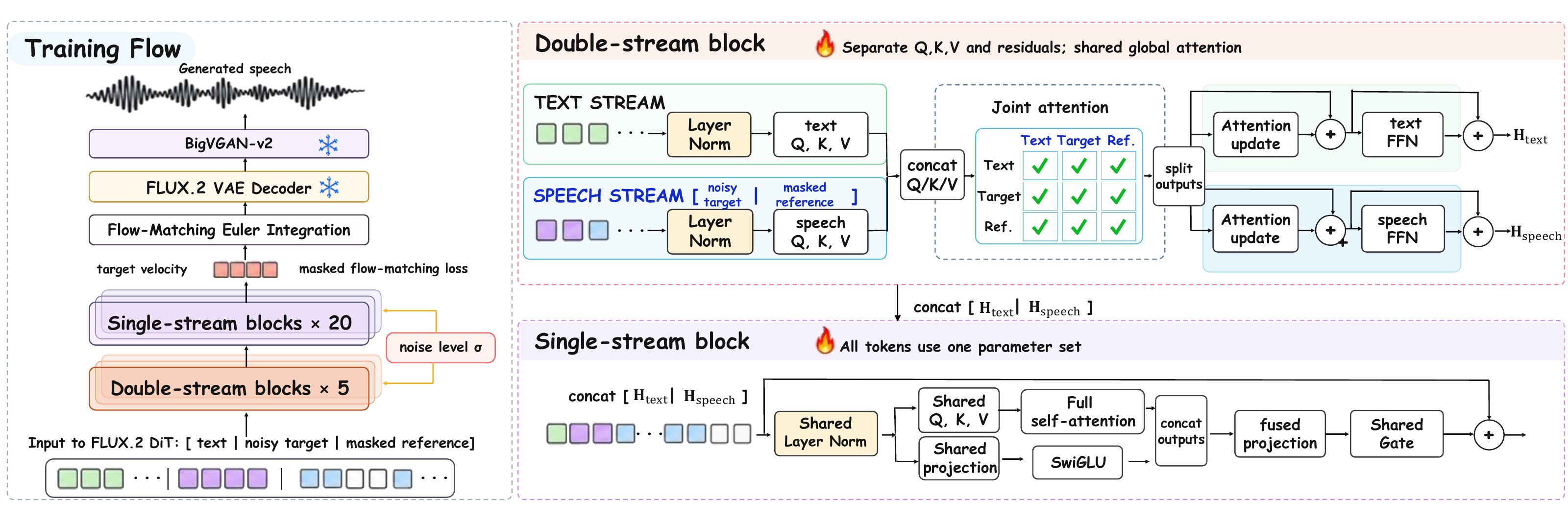}
    \caption{Architecture of ImageTTS-base. The framework adopts a Flow-Matching DiT backbone comprising double-stream blocks and single-stream blocks conditioned on timestep $\sigma$. High-fidelity speech is synthesized via Euler integration, a frozen FLUX.2 VAE decoder, and BigVGAN-v2.}
    \label{fig:overview}
    \vspace{-10pt}
\end{figure}
\subsection{Speech representation in the pretrained image latent space}
\label{sec:mel_preprocessing}

Speech is represented in the latent space of the frozen FLUX.2 image VAE.
Given a waveform, a single-channel log-mel spectrogram $\mathbf{M}_{\mathrm{mel}}\in\mathbb{R}^{F_{\mathrm{mel}}\times T_{\mathrm{mel}}}$ is first extracted, where $F_{\mathrm{mel}}=100$ denotes the number of mel-frequency bins and $T_{\mathrm{mel}}$ denotes the number of temporal frames. 
The frequency and temporal axes are interpreted as the height and width axes of an image, respectively.
The FLUX.2 VAE downsamples each spatial dimension by a factor of $8$, and is subsequently rearranged into $2\times2$ patches for DiT processing. Accordingly, both input spatial dimensions are required to be divisible by $16$. 
The mel-spectrogram is bilinearly resized from $100$ to $F$ frequency bins and right-padded to a multiple of $16$ frames using the utterance's minimum log-mel value to approximate silence. It is then linearly normalized to $[-1,1]$ using the training-set minimum and maximum log-mel values and replicated across three channels, yielding a pseudo-RGB input of shape $3\times F\times T$, where $T$ is the padded temporal length. The frozen VAE encoder maps this input to a latent representation, which is standardized using the pretrained VAE latent statistics.

\subsection{Multi-task conditional flow matching}
\label{sec:conditioned_fm}
Let $p_0$ denote the standard Gaussian distribution and $p_1$ denote the empirical distribution of target speech latents. For a target latent $\mathbf{z}^{\mathrm{tgt}}\sim p_1$ and Gaussian noise $\boldsymbol{\epsilon}\sim p_0$, we obtain a noise level $\sigma$ from the scheduler for a sampled training timestep.  We then construct the linear interpolation path and its corresponding target velocity as
\begin{equation}
\mathbf{z}_{\sigma}=(1-\sigma)\mathbf{z}^{\mathrm{tgt}}+\sigma\boldsymbol{\epsilon},\qquad
\mathbf{u}_{\sigma}=\frac{\partial\mathbf{z}_{\sigma}}{\partial\sigma}=\boldsymbol{\epsilon}-\mathbf{z}^{\mathrm{tgt}}.
\end{equation}
Here, $\sigma=1$ corresponds to Gaussian noise and $\sigma=0$ to the clean target latent.
Let $\mathbf{y}$ denote the text-token sequence, and let $\mathbf{z}^{\mathrm{ref}}$ denote an optional reference speech latent. Conditioned on these inputs, the FLUX.2 DiT predicts the velocity field
\begin{equation}
\widehat{\mathbf{v}}_{\sigma}=\mathbf{v}_{\theta}\left(\mathbf{z}_{\sigma},\sigma\mid \mathbf{y},\mathbf{z}^{\mathrm{ref}}\right).
\end{equation}
Let $\overline{\mathbf{m}}$ be a task-specific binary loss mask with the same shape as $\mathbf{z}^{\mathrm{tgt}}$. We train the model with the masked conditional flow-matching objective

\begin{equation}
\mathcal{L}_{\mathrm{CFM}}=\mathbb{E}_{\sigma,\mathbf{z}^{\mathrm{tgt}},\boldsymbol{\epsilon}}\left[\frac{\left\|\overline{\mathbf{m}}
\odot\left(\widehat{\mathbf{v}}_{\sigma}-\mathbf{u}_{\sigma}\right)\right\|_2^2}{\left\|\overline{\mathbf{m}}\right\|_1}\right],
\label{eq:conditional_fm}
\end{equation}
where $\odot$ denotes element-wise multiplication.
During inference, we initialize the target latent as $\mathbf{z}_{\sigma_0}\sim\mathcal{N}(\mathbf{0},\mathbf{I}),\sigma_0=1$. We then solve the conditional ordinary differential equation (ODE) along the scheduler's decreasing noise levels. The resulting $\mathbf{z}_{\sigma_N}$, with $\sigma_N=0$, is decoded as the generated speech latent. A numerical ODE solver evaluates the velocity field at a finite number of scheduler noise levels. This number of function evaluations (NFE) controls the trade-off between sampling accuracy and computational cost.

\paragraph{Zero-shot TTS.}
Zero-shot TTS is formulated as reference-conditioned latent infilling, implemented with masked flow matching. During training, for each utterance, we sample a contiguous temporal region with a random duration and starting position as the region to be generated. The binary temporal mask $\overline{\mathbf{m}}$, whose entries are set to one within the selected region and to zero elsewhere, is broadcast across the channel and frequency dimensions. The reference speech is provided through a masked reference-conditioning stream, $ \mathbf{z}^{\mathrm{ref}} = (\mathbf{1}-\overline{\mathbf{m}}) \odot \mathbf{z}^{\mathrm{tgt}}$, which retains the observed speech latent outside the selected region and is set to zero within the region to be generated. The text stream is encoded with the frozen Qwen3 text encoder. The noisy target latent $\mathbf{z}_\sigma$ and the masked reference-conditioning latent $\mathbf{z}^{\mathrm{ref}}$ are projected into  separate  token sequences and concatenated to form the speech stream. As illustrated in Fig.~\ref{fig:overview}, each double-stream block applies separate query, key, and value (Q, K, V) projections, feed-forward networks, and residual pathways to the text and speech streams. The resulting Q, K, and V tensors from the two streams are concatenated for joint bidirectional attention, enabling global interaction among the text, noisy target, and reference speech tokens. The resulting text and speech hidden states, $\mathbf{H}_{\mathrm{text}}$ and $\mathbf{H}_{\mathrm{speech}}$, are then concatenated along the sequence dimension and processed by the single-stream blocks through shared self-attention and feed-forward layers. The flow-matching loss in Eq.~\ref{eq:conditional_fm} is evaluated only within the selected region. During inference, the reference-conditioning stream contains the reference speech prefix and zeros over the continuation region, and is kept fixed throughout sampling. The target latent is updated over the full requested duration. After sampling, only the continuation suffix is retained and decoded as the output.

\paragraph{Instruction-guided TTS.}
Instruction-guided TTS generates the utterance from a transcript and a natural-language style instruction $\mathbf{i}$. We combine the two text inputs using the fixed prompt template 
\begin{center}
\texttt{Transcript: \{transcript\}}\\
\texttt{Style instruction: \{instruction\}}
\end{center}
and encode the resulting prompt with the same frozen Qwen3 text encoder. No reference-speech tokens are appended to the speech stream. We therefore set $\overline{\mathbf{m}}=\mathbf{1}$ in Eq.~\ref{eq:conditional_fm}, so that the flow-matching loss is
evaluated over the complete target speech latent.

\paragraph{Classifier-free guidance.}
We enable task-specific classifier-free guidance (CFG) through selective condition dropout during training. For zero-shot TTS, the transcript is replaced with an empty string with probability $p_{\mathrm{text}}$. In addition, the sampled temporal mask may cover the entire utterance. In this case, $\overline{\mathbf{m}}=\mathbf{1}$ and $\mathbf{z}^{\mathrm{ref}}=\mathbf{0}$, exposing the model to training examples without speech reference. At inference, we independently control the strengths of the text guidance and speech guidance as
\begin{equation}
\begin{aligned}
\widetilde{\mathbf{v}}_{\mathrm{ZS}}
={}&\mathbf{v}_{\mathbf{y},\mathbf{a}}
+(s_{\mathrm{t}}-1)
\left(
\mathbf{v}_{\mathbf{y},\mathbf{a}}
-\mathbf{v}_{\varnothing,\mathbf{a}}
\right)\\
&+(s_{\mathrm{s}}-1)
\left(
\mathbf{v}_{\mathbf{y},\mathbf{a}}
-\mathbf{v}_{\mathbf{y},\varnothing}
\right),
\end{aligned}
\label{eq:zs_cfg}
\end{equation}
where $\mathbf{y}$ denotes the text condition, $\mathbf{a}=\mathbf{z}^{\mathrm{ref}}_{\mathrm{inf}}$ denotes the reference
speech condition, and $\varnothing$ denotes the absence of the corresponding condition. The text CFG scale $s_{\mathrm{t}}$ strengthens conditioning on the transcript to improve linguistic content fidelity, whereas the speech CFG scale $s_{\mathrm{s}}$ strengthens conditioning on the reference speech to preserve speaker identity.

For instruction-guided TTS, classifier-free guidance is applied only to the style instruction. During training, the style instruction is independently replaced with an empty instruction with probability $p_{\mathrm{style}}$, while the transcript condition $\mathbf{y}$ is always retained. At inference, the guided velocity is computed as
\begin{equation}
\widetilde{\mathbf{v}}_{\mathrm{IT}}=\mathbf{v}_{\mathbf{y},\mathbf{i}}+(s_{\mathrm{i}}-1)\left(\mathbf{v}_{\mathbf{y},\mathbf{i}}-\mathbf{v}_{\mathbf{y},\varnothing}\right),
\label{eq:instruct_cfg}
\end{equation}
where $\mathbf{i}$ denotes the style instruction and $s_{\mathrm{i}}$ is the style CFG scale. 

\section{Analyzing transfer from image generation to speech}
\subsection{Speech reconstruction with a frozen image VAE}
To assess whether the frozen FLUX.2 VAE can serve as a latent interface for mel-spectrograms, we evaluate speech reconstruction after VAE encoding and decoding. We sample 100 utterances from each of the LibriTTS~(24kHz)~\citep{zen2019libritts} and LibriSpeech~(16kHz)~\citep{panayotov2015librispeech} datasets. The original waveforms serve as references. As a vocoder-only baseline, we synthesize waveforms directly from the mel-spectrograms using BigVGAN-v2~\citep{lee2023bigvgan}. For VAE reconstruction, we preprocess each mel-spectrogram and use bilinear interpolation to set its frequency-axis size to $F\in\{96,128,256,512,1024\}$. We then encode and decode it with the frozen VAE, invert the preprocessing, and synthesize a waveform with the same vocoder. Table~\ref{tab:vae-recon} reports the results, while definitions of the evaluation metrics are provided in Appendix~\ref{sec:add_metrics}.

For FLUX.2, reconstruction quality improves with increasing $F$ on both datasets, although the gains diminish at higher resolutions. Increasing $F$ from 256 to 512 improves PESQ by 0.40 on LibriTTS and 0.34 on LibriSpeech, whereas a further doubling to 1024 yields gains of only 0.12 and 0.09, respectively. Although $F=1024$ achieves slightly better reconstruction metrics, it doubles the number of latent positions relative to $F=512$. We therefore use $F=512$ in subsequent generation experiments to balance reconstruction quality and latent sequence length. At this resolution, STOI reaches 0.99 on LibriTTS and 0.98 on LibriSpeech, while SIM reaches 0.95 and 0.96, respectively. These values approach the vocoder-only baseline, although PESQ remains 0.57 lower on both datasets, indicating that this setting largely preserves intelligibility and speaker identity while leaving room for improvement in reconstruction quality.

We further compare FLUX.2 with the VAE used in Riffusion~\citep{Forsgren2022riffusion}, a spectrogram-oriented image-VAE baseline. We evaluate this VAE with both the original Riffusion preprocessing, denoted as \textit{Riffusion-o}, and preprocessing matched to ImageTTS, denoted as \textit{Riffusion-n}. Details of the preprocessing protocols are provided in Appendix~\ref{sec:riffusion_preprocessing}. Results show that the reconstruction quality of the Riffusion VAE is highly sensitive to preprocessing. Even with the matched preprocessing and the same frequency resolution of $F=512$, \textit{Riffusion-n} remains substantially worse than FLUX.2, indicating that the latter is more compatible with speech.

\begin{table*}[t]
\centering
\caption{
Speech reconstruction using different frozen image VAEs. The best results among the evaluated VAE configurations are shown in \textbf{bold}.
}
\label{tab:vae-recon}
\setlength{\tabcolsep}{3.0pt}
\renewcommand{\arraystretch}{0.94}

\resizebox{\textwidth}{!}{%
\begin{tabular}{l ccccc ccccc}
\toprule
\multirow{2}{*}{Method}
& \multicolumn{5}{c}{LibriTTS (24\,kHz)}
& \multicolumn{5}{c}{LibriSpeech (16\,kHz)} \\
\cmidrule(lr){2-6}
\cmidrule(lr){7-11}
& PESQ $\uparrow$
& STOI $\uparrow$
& SIM $\uparrow$
& DNSMOS $\uparrow$
& UTMOS $\uparrow$
& PESQ $\uparrow$
& STOI $\uparrow$
& SIM $\uparrow$
& DNSMOS $\uparrow$
& UTMOS $\uparrow$ \\
\midrule

Original waveform
& 4.64 & 1.00 & 1.00 & 3.74 & 4.01
& 4.64 & 1.00 & 1.00 & 3.93 & 4.13 \\

Vocoder only
& 4.43 & 1.00 & 0.98 & 3.75 & 3.99
& 4.36 & 0.99 & 0.98 & 3.89 & 3.99 \\

\midrule
\textit{Riffusion-o}
& 1.40 & 0.82 & 0.72 & 3.34 & 1.40
& 1.42 & 0.81 & 0.79 & 3.46 & 1.33 \\

\textit{Riffusion-n}, $F=512$
& 2.45 & 0.92 & 0.86 & 3.56 & 2.41
& 2.51 & 0.92 & 0.89 & 3.67 & 2.28 \\

\midrule
FLUX.2, $F=96$
& 2.67 & 0.94 & 0.90 & 3.50 & 2.66
& 2.61 & 0.94 & 0.91 & 3.57 & 2.51 \\

FLUX.2, $F=128$
& 3.01 & 0.96 & 0.91 & 3.58 & 2.88
& 2.98 & 0.95 & 0.93 & 3.69 & 2.76 \\

FLUX.2, $F=256$
& 3.46 & 0.97 & 0.93 & 3.66 & 3.24
& 3.45 & 0.97 & 0.95 & 3.79 & 3.16 \\

\rowcolor{PrivateRow}
FLUX.2, $F=512$
& 3.86 & \textbf{0.99} & 0.95 & \textbf{3.72} & 3.58
& 3.79 & 0.98 & \textbf{0.96} & 3.86 & 3.55 \\

FLUX.2, $F=1024$
& \textbf{3.98} & \textbf{0.99} & \textbf{0.96}
& \textbf{3.72} & \textbf{3.71}
& \textbf{3.88} & \textbf{0.99} & \textbf{0.96}
& \textbf{3.88} & \textbf{3.69} \\

\bottomrule
\end{tabular}
}
\vspace{-15pt}
\end{table*}

\subsection{Effect of image-pretrained initialization on speech adaptation}
\label{sec:image_prior_analysis}

\begin{wrapfigure}{r}{0.50\linewidth}
    \centering
    \includegraphics[width=\linewidth]{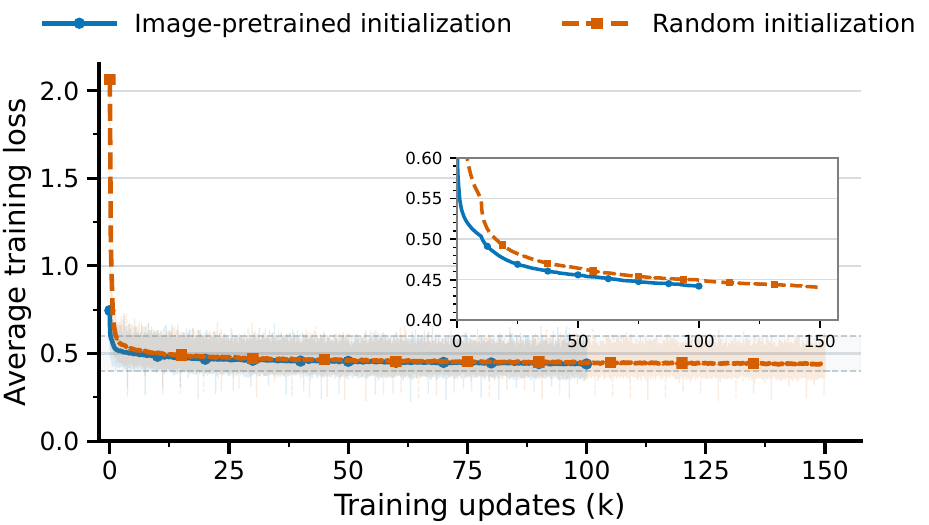}
    \caption{Training loss on LibriTTS for image-pretrained and randomly initialized DiTs operating in the frozen image-VAE latent space. Curves show the batch-averaged training loss at each update, smoothed with a 500-update moving-average window in the main panel. The inset enlarges the low-loss region between 0.4 and 0.6 and applies a 10,000-update smoothing window to highlight the long-term trend.}
    \label{fig:ablation_training_loss}
    \vspace{-10pt}
\end{wrapfigure}
This section investigates whether initializing the DiT with image-generation weights facilitates adaptation to speech. To isolate the effect of initialization, all compared models share the DiT architecture, image-VAE latent space, and optimization schedule, differing only in their initial weights. Models are trained for zero-shot TTS on LibriTTS, a widely used medium-scale TTS corpus containing approximately 585 hours of 24kHz read English speech from 2,456 speakers \citep{zen2019libritts}. Evaluation is conducted on 200 utterances from the official LibriSpeech test-clean split~\citep{panayotov2015librispeech}. Following~\citet{chen2025neural}, we adopt a speech-continuation protocol: the first three seconds of each utterance serve as the acoustic prompt, and the model generates the remainder conditioned on the transcript. Detailed configurations are provided in Appendix~\ref{sec:init_details}.

We first examine the convergence behavior of the two initializations. Figure~\ref{fig:ablation_training_loss} shows the complete training-loss curves in the main panel and enlarges the low-loss region between 0.4 and 0.6 in the inset. The image-pretrained DiT starts from a substantially lower loss and remains ahead throughout the overlapping training range. Notably, at 100k updates, it reaches approximately the same loss attained by the randomly initialized model only after 150k updates. This result indicates that image-pretrained initialization improves training efficiency, enabling the model to reach a comparable loss with fewer optimization steps and fewer sample exposures.   

\begin{figure}[t]
    \centering
    \includegraphics[width=0.95\linewidth]{
        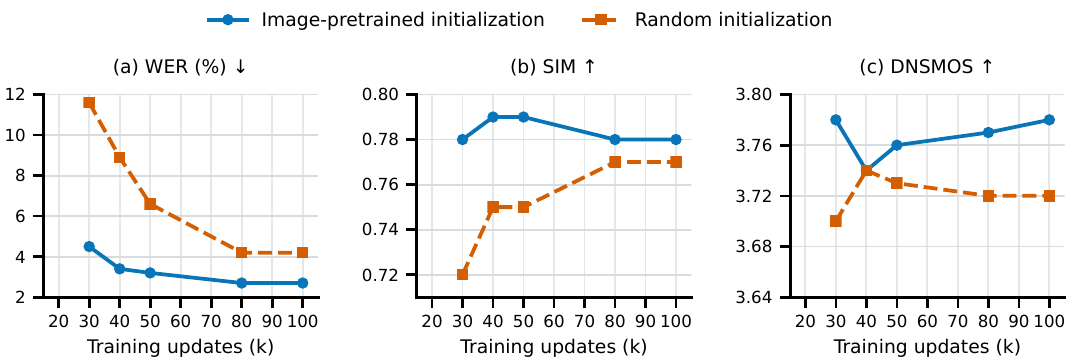
    }
    \caption{
        Generation metrics across training stages for image-pretrained and randomly initialized DiTs in the image-VAE latent space. Classifier-free guidance is fixed to 4. Markers denote evaluated checkpoints; connecting lines are visual guides. WER is reported as a percentage.
    }
    \label{fig:ablation_generation_metrics}
    \vspace{-10pt}
\end{figure}

Next, we evaluate the performance gap after sufficient training in Figure~\ref{fig:ablation_generation_metrics}. At the same training budget of 100k updates, the image-pretrained model outperforms the randomly initialized model on all three metrics. Across checkpoints, WER decreases steadily with training, while SIM remains consistently higher for the pretrained model, ranging from 0.78 to 0.79 compared with 0.75 to 0.77 for the randomly initialized model.
Even after the randomly initialized model is trained for 150k updates and reaches a comparable training loss, a substantial WER gap remains: 3.8\% for random initialization versus 2.7\% for image pretraining at 100k updates. 
These results suggest that the image prior provides a transferable generative prior that primarily benefits linguistic content generation while also modestly improving speaker preservation.

\subsection{Interaction between generation space and pretrained initialization}
\label{sec:ablation_generation_space}

We further compare flow-matching modeling in two representation spaces: modeling in the image-VAE latent space and the normalized mel-spectrograms. For direct mel modeling, we first normalize the mel-spectrogram using the global mean and standard deviation computed over the training set. We then patchify the normalized mel-spectrogram using patch sizes of $4$ and $8$ along the frequency and time dimensions, respectively, to shorten the sequence and reduce memory usage.  Next, a bias-free linear input adapter maps the 32-dimensional feature of each mel patch to the 128-dimensional input space expected by the pretrained DiT. The DiT output is then mapped back to a 32-dimensional patch representation through a symmetric linear adapter.

\begin{table}[h]
    \centering
    \caption{
    Effects of initialization and generation space at 100k training updates. Bold indicates the best result across all models.
    }
    \label{tab:ablation_initialization_space}

    \small
    \setlength{\tabcolsep}{5pt}
    \renewcommand{\arraystretch}{1.12}
    \resizebox{0.9\columnwidth}{!}{%
        \begin{tabular}{l ccc ccc}
            \toprule
            & \multicolumn{3}{c}{Image-pretrained}
            & \multicolumn{3}{c}{Random initialization} \\
            \cmidrule(lr){2-4}
            \cmidrule(lr){5-7}
            Generation space
            & WER~(\%) $\downarrow$
            & SIM $\uparrow$
            & DNSMOS $\uparrow$
            & WER~(\%) $\downarrow$
            & SIM $\uparrow$
            & DNSMOS $\uparrow$ \\
            \midrule
            Image-VAE latent
            & \textbf{2.7} & 0.78          & \textbf{3.78}
            & 4.2 & 0.77          & 3.72 \\
            Direct mel
            & 3.1          & \textbf{0.80} & 3.56
            & 4.2 & \textbf{0.80} & 3.51 \\
            \bottomrule
        \end{tabular}%
    }
\end{table}

Table~\ref{tab:ablation_initialization_space} reports all four combinations of the two initialization strategies and two modeling spaces. All models are evaluated at the same 100k-update checkpoint. Image-pretrained initialization reduces WER in both modeling spaces, indicating a consistent improvement in content accuracy. In contrast, the choice of modeling space involves a trade-off between speech quality and speaker similarity: across both initialization strategies, modeling in the image-VAE latent space achieves higher DNSMOS, whereas direct mel modeling yields higher speaker similarity. We hypothesize that the VAE latent representation provides a smoother and more structured prediction target. Although direct mel modeling preserves more speaker-related spectral information, it requires modeling longer sequences and finer-grained local variations, which may make spectral reconstruction less stable.

\begin{table*}[t]
  \centering
  \caption{
    Zero-shot TTS results on Seed-TTS test-en, Seed-TTS test-zh, and the internal English test set. MOS is reported as the mean score with a confidence interval (CI). The best and second-best results among models trained on the same source dataset (Emilia) are shown in \textbf{bold} and \underline{underlined}, respectively; ties receive the same marking. Missing results are denoted by --.
  }
  \vspace{-5pt}
  \label{tab:seedtts_results}
  \setlength{\tabcolsep}{3.3pt}
  \renewcommand{\arraystretch}{1.10}
  \resizebox{\textwidth}{!}{%
  \begin{tabular}{lr ccc ccc cc}
    \toprule
    \multirow{2}{*}{Method}
      & \multirow{2}{*}{Training hours}
      & \multicolumn{3}{c}{Seed-TTS test-en}
      & \multicolumn{3}{c}{Seed-TTS test-zh}
      & \multicolumn{2}{c}{Internal test-en} \\
    \cmidrule(lr){3-5}
    \cmidrule(lr){6-8}
    \cmidrule(lr){9-10}
      & &
      WER$\downarrow$ & SIM$\uparrow$ & MOS$\uparrow$
      & CER$\downarrow$ & SIM$\uparrow$ & MOS$\uparrow$
      & WER$\downarrow$ & SIM$\uparrow$ \\
    \midrule

    Ground Truth
      & --
      & 0.88 & 0.73 & $4.17 \pm 0.19$
      & 1.25 & 0.76 & $4.25 \pm 0.16$
      & -- & -- \\

    Reconstructed
      & --
      & 0.90 & 0.71 & --
      & 1.34 & 0.73 & --
      & -- & -- \\
    \midrule

    Qwen3-TTS \citep{hu2026qwen3}
      & $>$5M
      & 0.96 & 0.72 & $4.06 \pm 0.16$
      & 1.23 & 0.77 & $3.91 \pm 0.19$
      & 0.72 & 0.69 \\

    FireRedTTS3-Instruct \citep{shen2026fireredtts3}
      & $>$2.6M
      & 1.55 & 0.64 & $4.04 \pm 0.18$
      & 1.09 & 0.71 & $3.67 \pm 0.20$
      & 0.84 & 0.59 \\

    VoxCPM2 \citep{zhou2026voxcpm2}
      & $>$2M
      & 1.20 & 0.68 & $3.90 \pm 0.19$
      & 1.04 & 0.74 & $4.24 \pm 0.17$
      & 0.51 & 0.55 \\
    \midrule

    F5-TTS $^*$ \citep{chen2025f5}
      & 95K
      & \underline{1.17} & \underline{0.67} & $3.33 \pm 0.20$
      & \underline{1.70} & \underline{0.76} & \underline{$3.53 \pm 0.19$}
      & \underline{0.94} & \underline{0.64} \\

    MaskGCT $^*$ \citep{wang2025maskgct}
      & 100K
      & 2.33 & \textbf{0.71} & \underline{$3.95 \pm 0.19$}
      & 2.36 & \textbf{0.77} & $3.34 \pm 0.18$
      & 1.45 & \textbf{0.65} \\

    ImageTTS-base
      & 100K
      & \textbf{1.12} & 0.66 & \textbf{4.08 $\pm$ 0.18}
      & \textbf{1.68} & 0.72 & \textbf{4.01 $\pm$ 0.17}
      & \textbf{0.72} & \underline{0.64} \\

    \bottomrule
  \end{tabular}%
  }
  \vspace{2pt}
  \parbox{\textwidth}{%
    \footnotesize
    $^*$ denotes the primary baselines trained on comparable
    amounts of Emilia data to ImageTTS-base.
  }
  \vspace{-10pt}
\end{table*}

\subsection{Zero-shot TTS results}
Table~\ref{tab:seedtts_results} reports zero-shot TTS results on the Seed-TTS test sets and our internal English test set, which is described in Appendix~\ref{sec:internal_test_set}. We re-evaluate all baselines on all test sets using their official implementations and report results under a unified evaluation protocol. We group the baselines by training data. Systems marked with $^*$ are trained on approximately 100k hours of Emilia speech, comparable to ImageTTS-base, and serve as the primary baselines. The remaining systems are trained on substantially larger, non-public speech datasets and are included for reference.

Among systems trained on comparable amounts of Emilia data, ImageTTS-base achieves the lowest text error rate in every evaluation setting, indicating stable linguistic content generation across datasets. It also obtains the highest MOS on both the English and Chinese Seed-TTS test sets, reaching 4.08 and 4.01, respectively. Its speaker similarity is comparable to F5-TTS on the two English test sets, but lower than those of F5-TTS and MaskGCT on the Chinese set. Systems trained on substantially larger datasets achieve strong objective performance overall, with Qwen3-TTS leading most metrics and VoxCPM2 performing particularly well in Chinese content accuracy. Nevertheless, ImageTTS-base outperforms FireRedTTS3-Instruct on seven of the eight reported metrics and remains competitive in linguistic accuracy and perceived quality. Its main limitation is speaker similarity, especially on the Chinese Seed-TTS test set. We attribute this gap to limitations in the speaker-identity information preserved by the frozen image VAE. This interpretation is supported by the reconstruction results in Table~\ref{tab:vae-recon}, where image-VAE reconstruction achieves a speaker similarity of 0.95, compared with 0.98 for direct mel-spectrogram modeling followed by the same vocoder.

\begin{figure*}[t]
    \centering
    \includegraphics[width=\textwidth,
    trim=0 0.3cm 0 0.2cm,
        clip
        ]{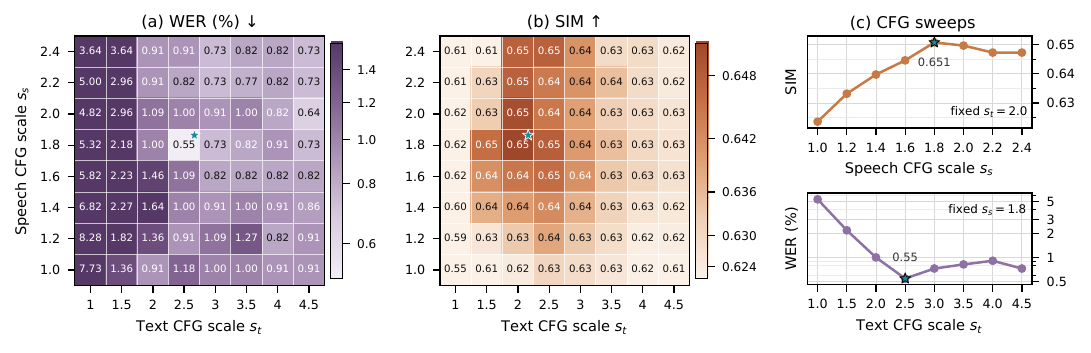}
     \vspace{-15pt}
    \caption{
    Effects of text and speech CFG scales. Panels (a) and (b) report WER and SIM, respectively, over a grid search of the text and speech CFG scales $s_t$ and $s_s$. Stars indicate the configurations with the lowest WER and highest SIM, respectively. Panel (c) plots WER as a function of $s_t$ with $s_s=1.8$, and SIM as a function of $s_s$ with $s_t=2.0$.}
    \label{fig:dual_cfg_ablation}
     \vspace{-10pt}
\end{figure*}

\subsubsection{Effect of dual classifier-free guidance}
We perform a grid search over the text and speech CFG scales on the first 100 examples of the Seed-TTS test-en set to examine their effects on linguistic accuracy and speaker similarity. For ease of interpretation, we first fix the text CFG scale at $s_t=2.0$ and vary the speech CFG scale. As shown in the upper plot of Fig.~\ref{fig:dual_cfg_ablation}~(c), speaker similarity increases with $s_s$ up to 1.8, reaching a maximum of 0.651, and then remains nearly unchanged or slightly decreases. We next fix $s_s=1.8$ and vary $s_t$. As shown in the lower plot, WER decreases from 5.32\% at $s_t=1.0$ to a minimum of 0.55\% at $s_t=2.5$, after which it increases again. The heatmaps in Fig.~\ref{fig:dual_cfg_ablation}~(a,b) further show that excessively strong text guidance can slightly reduce speaker similarity, whereas increasing the speech guidance beyond $s_s=1.8$ provides little additional benefit. These results suggest that the two CFG scales should be tuned jointly; in this sweep, $(s_t,s_s)=(2.5,1.8)$ provides a favorable balance between linguistic accuracy and speaker similarity.
\begin{table*}[h]
    \centering
    \caption{
    Instruction-guided TTS results on the CapSpeech test set.
    The best and second-best results among models trained
    on CapSpeech instruction data are shown in \textbf{bold}
    and \underline{underlined}, respectively.
    N/A indicates that the amount of instruction data is not available.
    }
    \vspace{-5pt}
    \label{tab:instruct_tts_main}
    \setlength{\tabcolsep}{3.0pt}
    \renewcommand{\arraystretch}{1.10}

    \resizebox{\textwidth}{!}{%
    \begin{tabular}{l c c c c c c c c}
        \toprule
        Method
        & Instruction data
        & Age $\uparrow$
        & Gender $\uparrow$
        & Pitch $\uparrow$
        & Expressiveness $\uparrow$
        & Style-ACC $\uparrow$
        & CLSP-Score $\uparrow$
        & WER $\downarrow$ \\
        \midrule

        PromptTTS
        & $\sim$0.3K h
        & 0.48 & 0.59 & 0.12 & 0.09
        & 0.32 & 0.45 & 6.3 \\

        VoxInstruct
        & 2.6K h
        & 0.76 & 0.99 & 0.23 & 0.41
        & 0.60 & 0.50 & 8.6 \\

        FireRedTTS3-Instruct
        & 330K h
        & 0.56 & 0.92 & 0.22 & 0.47
        & 0.54 & 0.48 & 6.7 \\

        Qwen3-TTS-VD
        & N/A
        & 0.66 & 0.90 & 0.18 & 0.42
        & 0.54 & 0.50 & 6.1 \\

        VoxCPM2
        & $\sim$10K h
        & 0.87 & 0.97 & 0.46 & 0.48
        & 0.69 & 0.48 & 5.7 \\
        \midrule

        CapTTS-AR
        & 33.6K h
        & 0.55 & \textbf{0.98} & 0.18 & 0.36
        & 0.52 & \underline{0.45} & 11.9 \\

        CapTTS-NAR
        & 33.6K h
        & \underline{0.79} & \underline{0.95}
        & \underline{0.36} & \underline{0.41}
        & \underline{0.62} & \textbf{0.49} & \underline{10.8} \\

        ImageTTS-instruct
        & 9K h
        & \textbf{0.88} & 0.89
        & \textbf{0.39} & \textbf{0.49}
        & \textbf{0.66} & \textbf{0.49} & \textbf{8.8} \\

        \bottomrule
    \end{tabular}%
    }
    \vspace{-10pt}
\end{table*}

\subsection{Instruction-guided TTS results}
Having established ImageTTS-base for zero-shot TTS, we next extend the framework to instruction-guided TTS to investigate natural-language control over speaking style. In the zero-shot setting, Qwen3 receives only the transcript and encodes the linguistic content required for speech generation. Instruction-guided TTS therefore requires Qwen3 to interpret the semantics of the instruction, while the DiT must learn to ground these semantics in the speech generation space. We initialize ImageTTS-instruct from ImageTTS-base and continue training it on paired speech--instruction data from CapSpeech. Implementation details are provided in Appendix~\ref{sec:scaling_set_ups}. 
Table~\ref{tab:instruct_tts_main} compares instruction-guided speech synthesis methods on a 200-example subset of the CapSpeech test set. Detailed metric definitions are provided in Appendix~\ref{sec:add_metrics}.

Among models trained on CapSpeech instruction data, ImageTTS-instruct achieves the best performance on nearly all metrics despite using only 9K hours of instruction data, as some subsets are excluded from training, compared with 33.6K hours for the CapTTS baselines. When further compared with systems trained on other instruction datasets, ImageTTS-instruct remains competitive on the two overall instruction-following metrics, achieving a Style-ACC of 0.66 and a CLSP-Score of 0.49, with only a small gap from VoxCPM2. These results suggest that the frozen Qwen3 text encoder provides effective semantic representations of natural-language instructions, while the generative backbone can learn their relation with the speech generation space. With larger and higher-quality instruction data, this modeling paradigm may offer further potential.

\section{Conclusion}
We presented ImageTTS, a framework that repurposes the FLUX.2 text-to-image pipeline for speech synthesis by modeling pseudo-RGB mel-spectrograms in the latent space of a frozen image VAE. In controlled LibriTTS experiments, the frozen FLUX.2 VAE preserves speech intelligibility and speaker identity, achieving 0.99 STOI and 0.95 speaker similarity, while image-pretrained DiT initialization substantially improves adaptation efficiency and reduces WER relative to matched training from scratch. When scaled to 100K hours of speech and applied to both zero-shot and instruction-guided TTS, ImageTTS achieves 1.12\% WER on the Seed-TTS test-en set and a Style-ACC of 0.66. These results suggest that pretrained image-generation models can transfer beyond vision, while also highlighting the importance of the chosen latent space and adaptation strategy. A natural direction for future work is to adapt the image VAE to speech to improve speaker preservation and perceptual speech quality.

\section*{AI Use Statement}
We used generative AI tools to assist with manuscript language editing and to obtain feedback on the presentation and interpretation of experimental results and possible follow-up analyses. The authors independently checked the resulting text and interpretations against the experimental results and relevant literature. The authors are responsible for all claims and content in this paper.

\section*{Reproducibility Statement}

We provide training and inference scripts, together with the reference speech prompts and target transcripts for our internal English test set, in the supplementary material for review. Experimental settings and evaluation protocols are described in the main paper and appendix. Upon acceptance, we will publicly release the complete codebase and trained model checkpoints.Here is the anonymous demo page of ImageTTS: \url{https://paperdemos.github.io/ImageTTS-demos/.}

\bibliography{iclr2027_conference}
\bibliographystyle{iclr2027_conference}

\appendix
\section{Preliminaries: the FLUX.2-Klein-4B pipeline}
FLUX.2-klein-base-4B is a conditional flow-based generative model composed of three principal components: a Qwen3 text encoder, a variational autoencoder (VAE), and a $3.9$B-parameter multimodal Transformer. The text encoder converts natural-language prompts into contextual representations, while the VAE maps images between the pixel space and a compact continuous latent space. The FLUX.2 Transformer jointly processes the textual and visual representations and predicts the velocity field used to generate latent samples through flow integration.

\paragraph{Text Encoder.}
The pipeline uses a pretrained Qwen3 language model to encode the input prompt. Given a tokenized text sequence $\mathbf{y}$ of length $L$, the language model produces a hierarchy of contextual hidden states. FLUX.2-Klein-4B extracts representations from three intermediate layers, indexed by $\{9,18,27\}$, and concatenates them along the feature dimension:
\begin{equation}
    \mathbf{c}
    =
    \operatorname{Concat}
    \left(
        \mathbf{h}^{(9)},
        \mathbf{h}^{(18)},
        \mathbf{h}^{(27)}
    \right)
    \in \mathbb{R}^{L\times7680},
\end{equation}
where
$\mathbf{h}^{(\ell)}\in\mathbb{R}^{L\times2560}$
denotes the hidden representation from the $\ell$-th layer. A learned context projection subsequently maps the concatenated representation to the Transformer hidden dimension:
\begin{equation}
    \widetilde{\mathbf{c}}
    =
    W_{\mathrm{ctx}}\mathbf{c}
    \in \mathbb{R}^{L\times3072}.
\end{equation}
The use of multiple intermediate layers provides the generative Transformer with contextual features at different levels of abstraction. The text encoder remains frozen during the training of the generative model.

\paragraph{Variational Autoencoder.}
FLUX.2 operates in a compressed continuous latent space defined by a pretrained VAE. Given an input image
$\mathbf{I}\in\mathbb{R}^{3\times H\times W}$,
the VAE encoder produces a spatial latent representation
\begin{equation}
    \mathbf{z}
    =
    \mathcal{E}_{\mathrm{VAE}}(\mathbf{I})
    \in
    \mathbb{R}^{32\times H/8\times W/8}.
\end{equation}
A subsequent $2\times2$ patchification operation folds each spatial neighborhood into the channel dimension:
\begin{equation}
    \mathbf{z}_{p}
    \in
    \mathbb{R}^{128\times H/16\times W/16}.
\end{equation}
Flattening the spatial dimensions yields a sequence of continuous visual tokens,
\begin{equation}
    \mathbf{x}
    \in
    \mathbb{R}^{N\times128},
    \qquad
    N=\frac{HW}{256}.
\end{equation}
The latent tokens are standardized using the VAE latent statistics and projected into the Transformer hidden space:
\begin{equation}
    \widetilde{\mathbf{x}}
    =
    W_{\mathrm{in}}\mathbf{x}
    \in
    \mathbb{R}^{N\times3072}.
\end{equation}
After generation, the inverse patchification operation reconstructs the spatial latent tensor, which is mapped back to the pixel space by the VAE decoder
$\mathcal{D}_{\mathrm{VAE}}$.

\paragraph{FLUX.2 Transformer.}
The FLUX.2-klein-base-4B denoiser contains approximately $3.9$B parameters and uses a hidden dimension of $3072$. Its backbone consists of five double-stream Transformer blocks followed by twenty single-stream Transformer blocks. Each block contains $24$ attention heads with a head dimension of $128$ and a feed-forward expansion ratio of $3$.

In the double-stream stage, textual and visual tokens retain separate residual streams. Each modality is processed by its own normalization layers, query--key--value projections, output projections, and feed-forward networks. Cross-modal communication is introduced by concatenating the text and visual queries, keys, and values before computing attention:
\begin{equation}
\begin{aligned}
    \mathbf{Q} &=
    [\mathbf{Q}_{\mathrm{text}};
     \mathbf{Q}_{\mathrm{vis}}],\\
    \mathbf{K} &=
    [\mathbf{K}_{\mathrm{text}};
     \mathbf{K}_{\mathrm{vis}}],\\
    \mathbf{V} &=
    [\mathbf{V}_{\mathrm{text}};
     \mathbf{V}_{\mathrm{vis}}],
\end{aligned}
\end{equation}
followed by
\begin{equation}
    \operatorname{Attn}(\mathbf{Q},\mathbf{K},\mathbf{V})
    =
    \operatorname{softmax}
    \left(
        \frac{\mathbf{Q}\mathbf{K}^{\top}}{\sqrt{d}}
    \right)\mathbf{V}.
\end{equation}
This formulation allows text and visual tokens to exchange information through bidirectional joint attention while preserving modality-specific transformations in the first five blocks.

After the double-stream stage, the updated representations are concatenated into a unified sequence:
\begin{equation}
    \mathbf{s}
    =
    [\widetilde{\mathbf{c}};
     \widetilde{\mathbf{x}}],
\end{equation}
which is processed by twenty single-stream Transformer blocks. These blocks apply shared attention and feed-forward transformations to all tokens, enabling fully fused multimodal processing. The visual portion of the final sequence is projected back to the latent dimensionality to produce the model output.

FLUX.2 uses multi-axis rotary positional embeddings to represent the heterogeneous structures of the two modalities. Text tokens are indexed along a one-dimensional sequence axis, whereas visual tokens are assigned two-dimensional spatial coordinates. The rotary embedding dimensions are distributed across four positional axes, with each axis occupying $32$ dimensions of the $128$-dimensional attention head. The flow timestep is encoded separately through sinusoidal embeddings and injected into each Transformer block through adaptive normalization.

\section{Baselines}
\textbf{F5-TTS}~\citep{chen2025f5}.
A fully non-autoregressive mel-spectrogram generator based on flow matching and a Diffusion Transformer, trained on approximately 95k hours of Emilia speech.

\textbf{MaskGCT}~\citep{wang2025maskgct}.
A two-stage masked codec Transformer that generates semantic and acoustic tokens non-autoregressively and is trained on the same 100k-hour bilingual Emilia corpus as ImageTTS-base.

\textbf{Qwen3-TTS}~\citep{hu2026qwen3}.
A large-scale dual-track language-model-based TTS system for which we use the Base checkpoint for zero-shot voice cloning and the separate VoiceDesign checkpoint, denoted as Qwen3-TTS-VD, for instruction-guided synthesis.

\textbf{FireRedTTS3-Instruct}~\citep{shen2026fireredtts3}.
A continuous autoregressive model with semantically enriched speech
representations, of
which FireRedTTS3-Instruct is evaluated in our experiments.

\textbf{VoxCPM2}~\citep{zhou2026voxcpm2}.
A hierarchical autoregressive-diffusion model that supports multilingual voice cloning and natural-language voice control within a single unified checkpoint.

\textbf{PromptTTS}~\citep{leng2024prompttts}.
A natural-language-guided TTS system that separately encodes content and style descriptions to control the generated speech.

\textbf{VoxInstruct}~\citep{zhou2024voxinstruct}.
A multilingual codec language model that directly generates expressive speech from free-form human instructions.

\textbf{CapTTS-AR and CapTTS-NAR}~\citep{wang2025capspeech}.
The autoregressive and non-autoregressive baselines released with CapSpeech, trained on the same style-captioned data as ImageTTS-instruct. The baselines released with CapSpeech comprise an autoregressive model (CapTTS-AR), initialized from the publicly available Parler-TTS Mini v1 checkpoint~\footnote{https://huggingface.co/parler-tts/parler-tts-mini-v1}, and a non-autoregressive model (CapTTS-NAR) based on the F5-TTS architecture~\citep{chen2025f5}.

\section{Evaluation metrics}
\label{sec:add_metrics}
\subsection{General metrics for TTS}
Perceptual Evaluation of Speech Quality (\textbf{PESQ})~\citep{rix2001perceptual}: measures the perceptual quality of the synthesized speech relative to the reference speech. 

Short-Time Objective Intelligibility (\textbf{STOI})~\citep{taal2010short}: evaluates speech intelligibility by comparing the temporal structures of the synthesized and reference speech. 

Speaker Similarity (\textbf{SIM})~\citep{chen2022wavlm}: measures speaker identity preservation as the cosine similarity between the WavLM-TDNN embeddings of the synthesized speech and the reference speech.

Deep Noise Suppression Mean Opinion Score (\textbf{DNSMOS})~\citep{reddy2021dnsmos}: estimates the perceptual quality of synthesized speech.

UTokyo-SaruLab Mean Opinion Score (\textbf{UTMOS})~\citep{saeki22c_interspeech}: predicts the mean opinion score of synthesized speech.

Word Error Rate (\textbf{WER}): measures recognition errors at the word level obtained using a Conformer-Transducer model~\citep{gulati2020conformer}. 

Character Error Rate (\textbf{CER}): measures recognition errors at the character level and is computed using Paraformer-zh~\citep{gao2022paraformer} for Chinese speech.

\subsection{Subjective metric for TTS}
We conduct a subjective mean opinion score (MOS) evaluation to assess the perceptual quality of zero-shot TTS samples on a 1-5 scale. For each system, we evaluate 30 generated utterances, including 15 English and 15 Chinese samples, with each utterance independently rated by 10 listeners. We report the average MOS with a 95\% confidence interval (CI).  

\subsection{Metrics for Instruction-guided TTS}
We report attribute-level classification accuracy for age, gender, pitch, and expressiveness, together with the aggregate Style-ACC following~\cite{wang2025capspeech}. We omit duration accuracy because non-autoregressive systems rely on externally estimated durations, making this metric partly reflect the duration estimation procedure rather than the model's instruction-following ability.

\textbf{Age \& Gender accuracy:} Evaluated using Wav2Vec 2.0 backbone~\footnote{https://huggingface.co/audeering/wav2vec2-large-robust-6-ft-age-gender}. For age, the model outputs a continuous value $\text{logits}_{\text{age}} \times 100$, which is discretized into five categories: \textit{child} ($\le 12$), \textit{teenager} ($\le 19$), \textit{young adult} ($\le 39$), \textit{middle-aged adult} ($\le 64$), and \textit{elderly} ($> 64$). For gender, the prediction is determined by taking the argmax over the binary logits (\texttt{female} vs. \texttt{male}), disregarding the intermediate child class.

\textbf{Pitch:} Extracted via the PENN~\footnote{https://github.com/interactiveaudiolab/penn}. We compute the mean frame-level $F_0$ and discretize it into 7 pitch tiers using calibrated, gender-dependent threshold tables following~\cite{wang2025capspeech}.

\textbf{Expressiveness:} Expressiveness is proxied by the frame-level pitch standard deviation estimated by the same PENN. This metric reflects speech melody and dynamic range, which is mapped into 5 categorical bins ranging from \textit{very monotone} to \textit{very expressive and animated} according to pre-calibrated monotony thresholds following~\cite{wang2025capspeech}.

\textbf{Style-ACC:}
Style-ACC is a sample-level average classification accuracy across the evaluated style attributes.

\textbf{CLSP-Score:}
Cross-modal semantic consistency between the instruction prompt and the generated speech waveform is measured using the CLSP model~\citep{yang2026clsp}. For each sample, normalized text embedding and audio embedding are extracted. The global CLSP-Score is reported as the average cosine similarity over the test set.
\section{Riffusion preprocessing}
\label{sec:riffusion_preprocessing}
We first compute a 100-bin Mel spectrogram in the linear-amplitude domain from the input waveform. To satisfy the input-size requirements of the Riffusion VAE, we pad the frequency and temporal dimensions to multiples of 32, expanding the frequency dimension from 100 to 128 bins and padding the temporal dimension accordingly. The padding value is set to the minimum spectrogram amplitude. We then apply Riffusion’s image preprocessing, including maximum-value normalization, a power transform with $\text{power}=0.25$, image-intensity inversion, and vertical flipping, before converting the result to RGB format for VAE encoding. After VAE reconstruction, we apply the corresponding inverse transformations, remove the padded regions, and recover the 100-bin log-Mel representation required by BigVGAN.

\section{Image-pretrained initialization for speech adaptation}
\label{sec:init_details}
This appendix details the initialization study in Section~\ref{sec:image_prior_analysis}. Both models use the same architecture, data, objective, and training procedure; only the DiT initialization differs. The image-pretrained model loads DiT parameters from FLUX.2-Klein-base-4B, whereas the randomly initialized model uses the same configuration with default initialization and a fixed random seed. The image VAE and Qwen3 text encoder remain frozen. We train the models on 16 NVIDIA A100 GPUs with 40 GB memory using BF16 mixed precision. Optimization uses AdamW with a constant learning rate of $1\times10^{-4}$ and an effective batch size of 128. The image-pretrained model is trained for 100k updates, while the randomly initialized model is trained for 150k updates to reach a comparable training loss. For convergence analysis, we evaluate checkpoints at 30k, 40k, 50k, 80k, and 100k updates on the same fixed set of 200 test utterances. The text and speech CFG scales are fixed to $s_{\mathrm{t}}=4$ and
$s_{\mathrm{s}}=1$ for all checkpoints to ensure a consistent comparison. We additionally evaluate the 150k-update checkpoint of the randomly initialized model. At this checkpoint, it achieves a WER of $3.8\%$, a speaker similarity of $0.76$, and a DNSMOS score of $3.75$.

\section{Experimental setup for scaling task}
\label{sec:scaling_set_ups}

\textbf{Datasets.} For zero-shot TTS, we train on the English and Chinese subsets of Emilia, which contain approximately 100k hours of publicly available speech collected from diverse in-the-wild sources. We evaluate on Seed-TTS \textit{test-en} and \textit{test-zh}, using 1,088 English and 2,020 Chinese samples, respectively. For instruction-guided TTS, we use the publicly released CapSpeech dataset, which pairs speech with natural-language descriptions of speaker characteristics and speaking styles. We use 3.47 million speech-caption pairs, totaling approximately 9K hours of speech, from the MLS-English, Emilia-English, and Common Voice portions of \textit{train-PT-CapTTS}. Our training data comes from the same collection as that used by CapTTS, but differs in size because some datasets were unavailable and some audio paths could not be reliably resolved. All log-mel spectrograms are extracted with 100 mel filterbanks using a Short-Time Fourier Transform (STFT) configured with an FFT size of 1024, a Hann window of length 1024, and a hop length of 256.

\textbf{Training.} 
We initialize ImageTTS from FLUX.2-klein-base-4B and fine-tune all DiT parameters on 64 NVIDIA A100 GPUs (40 GB each) using BF16 mixed precision. We use AdamW with a constant learning rate of $2\times10^{-4}$ and an effective batch size of 256, corresponding to approximately 0.7 hours of speech. The zero-shot TTS model is trained for 180k updates and is denoted as ImageTTS-base. For instruction-guided TTS, we further train the model for 50k updates, mixing instruction-guided and zero-shot TTS samples at a 1:1 ratio by sample count. The resulting checkpoint, denoted as ImageTTS-instruct, supports both zero-shot and instruction-guided TTS. We independently replace the transcript and style instruction with an empty string with probabilities $p_{\mathrm{text}}=0.1$ and $p_{\mathrm{style}}=0.1$, respectively.

\begin{table}[h]
\centering
\caption{
Pseudo-code for zero-shot TTS inference.
}
\label{tab:zs_sampling}
\small
\renewcommand{\arraystretch}{1.15}
\begin{tabular}{@{}p{0.06\linewidth}p{0.88\linewidth}@{}}
\hline
& \textbf{Input:} a reference speech latent
$\mathbf{z}^{\mathrm{ref}}$, text condition $\mathbf{y}$,
and the requested total target width $L_{\mathrm{req}}$,
which covers time columns of both the reference speech $\ell_{\mathrm{ref}}$ and the continuation target. \\

1 &
Initialize the complete target latent with standard Gaussian noise:
$\mathbf{z}_{\sigma_0} \sim \mathcal{N}(\mathbf{0},\mathbf{I})$. \\

2 &
Construct the full-size reference conditioning stream:
$\mathbf{c}\leftarrow\mathbf{0}$,\quad $\mathbf{c}[..., :\ell_{\mathrm{ref}}]\leftarrow\mathbf{z}^{\mathrm{ref}}$.\\

3 & 
\textbf{For} $k=0,\ldots,NFE-1$: \\

4 &
\quad Predict the full conditional velocity:
$\mathbf{v}^{\mathrm{full}}_k \leftarrow v_{\theta}\!\left( \operatorname{Concat}(\mathbf{z}_k,\mathbf{c}), \sigma_k \mid \mathbf{y}
\right)$. \\

5 &
\quad If text guidance is enabled, 
$
\mathbf{v}_k
\leftarrow
\mathbf{v}^{\mathrm{full}}_k
+
(s_t-1)
\left(
\mathbf{v}^{\mathrm{full}}_k
-
\mathbf{v}^{\mathrm{no\text{-}text}}_k
\right).
$\\

6 &
\quad If speech guidance is enabled, 
$\mathbf{v}_k
\leftarrow
\mathbf{v}_k
+
(s_s-1)
\left(
\mathbf{v}^{\mathrm{full}}_k
-
\mathbf{v}^{\mathrm{no\text{-}speech}}_k
\right).$
 \\

7 &
\quad Update using one step of the Euler ODE solver:

\quad$\mathbf{z}_{k+1}\leftarrow\mathbf{z}_k+(\sigma_{k+1}-\sigma_k)\mathbf{v}_k,$ where $\{\sigma_k\}_{k=0}^{NFE}$ is the noise level. \\

8 &
\textbf{End for}. \\

9 &
Recover the waveform by inverting preprocessing pipeline, and applying BigVGAN-v2. \\

\hline
\end{tabular}
\end{table}

\textbf{Inference.} We encode the reference speech into a standardized latent using the preprocessing pipeline and frozen VAE described in Section~\ref{sec:mel_preprocessing}. We then construct a conditioning tensor $\mathbf{c}$ with the same shape as the complete target latent $\mathbf{z_0}$, place the reference latent in its temporal prefix, and keep $\mathbf{c}$ fixed throughout sampling. The target latent is initialized with Gaussian noise and iteratively updated using an Euler ODE solver with dual classifier-free guidance and $NFE=50$. All of its positions, including the prefix according to the reference speech, are updated during sampling. Finally, we invert the latent and mel transformations, decode the mel-spectrogram with the frozen VAE and synthesize the waveform using BigVGAN-v2.

\section{Internal English test set}
\label{sec:internal_test_set}
We collect 20 reference utterances from 11 speakers across different countries. The utterances achieve an average UTMOS score of 4.15 and are manually screened for clear and natural pronunciation. For each reference utterance, GPT-5 generates five text prompts covering diverse sentence structures and levels of linguistic complexity. The internal evaluation set will be released after the anonymous review period.

\section{Duration estimation}
For zero-shot TTS, we remove leading and trailing silence from the reference speech and estimate the speaking rate by dividing the number of valid Chinese characters or English phonemes by the speech duration. To avoid distortion from extreme speaking rates, we constrain the rate: Chinese rates below 3 characters/s are set to 3, while rates above 6 characters/s are reduced to 6 characters/s.

For instruction-guided TTS, we additionally estimate the target duration from the speaking-rate description in the instruction. We use base rates of 5 Chinese characters/s and 14 English phonemes/s, and scale them by factors of (0.6), (0.8), (1.0), (1.1), and (1.2) for very slow, slow, moderate, fast, and very fast speech, respectively.  Because the duration is explicitly determined by this procedure rather than predicted by the model, we exclude duration control from the style-control accuracy metric for instruction-guided TTS.  

\section{An analysis of text encoder}
\label{sec:analysis}

\begin{CJK*}{UTF8}{gbsn}

\begin{table}[t]
    \centering
    \caption{
    Pronunciation sensitivity to punctuation-induced changes in
    text conditioning. Each case pairs the original input (O) with its
    punctuation-modified version (M). Metrics are sentence-level rates
    in percent; the displayed text and Pinyin cover only the target spans.
    }
    \label{tab:punctuation_cases}
    \scriptsize
    \setlength{\tabcolsep}{3pt}
    \renewcommand{\arraystretch}{1.12}
    \begin{tabularx}{\linewidth}{@{}l
        >{\raggedright\arraybackslash}X
        >{\raggedright\arraybackslash}p{0.29\linewidth}
        rr@{}}
        \toprule
        Case & Input span / Qwen token IDs
        & ASR-derived Pinyin & CER & PSER \\
        \midrule

        ZH-01 O
        & 贡献力量\newline
          \texttt{[118981]}
        & chuang4 jin4 ming2 liang4
        & 22.22 & 16.67 \\
        ZH-01 M
        & 贡献，力量\newline
          \texttt{[102007, 3837, 101102]}
        & gong4 xian4 li4 liang4
        & \textbf{0.00} & \textbf{0.00} \\
        \addlinespace

        ZH-02 O
        & 武汉麻将以\newline
          \texttt{[102051, 99965, 108794]}
        & wu3 han4 ma2 $\varnothing$ yi3
        & 26.67 & 6.67 \\
        ZH-02 M
        & 武汉麻将，以\newline
          \texttt{[102051, 112492, 3837, 23031]}
        & wu3 han4 ma2 jiang1 yi3
        & \textbf{6.67} & \textbf{0.00} \\
        \addlinespace

        ZH-03 O
        & 市委常委，常务副市长\newline
          \texttt{[113419, 3837, 109295, 102975]}
        & shi4 ye4 zhang3 wei3 chang2 fu4 bu4 shi4 zhang3
        & 35.71 & 28.57 \\
        ZH-03 M
        & 市委，常委，常务，副市长\newline
          \texttt{[102490, 3837, 101785, 3837,
          109294, 3837, 111251]}
        & shi4 wei3 chang2 wei3 chang2 wu4 fu4 shi4 zhang3
        & \textbf{0.00} & \textbf{0.00} \\
        \addlinespace

        ZH-04 O
        & 黑龙江省\newline
          \texttt{[111459]}
        & hei1 dong1 jiang1 sheng3
        & 5.26 & 5.26 \\
        ZH-04 M
        & 黑龙江，省\newline
          \texttt{[102951, 3837, 65770]}
        & hei1 long2 jiang1 sheng3
        & \textbf{0.00} & \textbf{0.00} \\
        \midrule

        \multicolumn{3}{@{}l}{Aggregate (O)}
        & 21.21 & 13.64 \\
        \multicolumn{3}{@{}l}{Aggregate (M)}
        & \textbf{1.52} & \textbf{0.00} \\
        \bottomrule
    \end{tabularx}
    \par\smallskip
    \begin{minipage}{\linewidth}
        \scriptsize
        Pinyin is automatically converted from ASR text and is not an
        independent phonetic transcription. Automatic conversion may
        misresolve polyphonic characters.
        $\varnothing$ denotes a missing syllable.
        In ZH-02, the residual CER error is the homophonic substitution
        ``赖''/``癞'' (\texttt{lai4}).
    \end{minipage}
\end{table}
We examine four selected Chinese failure cases by inserting commas while
preserving the original Chinese characters and their order.
Table~\ref{tab:punctuation_cases} reports the resulting tokenization changes
and recognition errors. Besides CER, we measure a tone-aware Pinyin
syllable error rate (PSER) by converting the reference and ASR transcription
into syllables with tone numbers and computing their Levenshtein distance,
normalized by the reference syllable count. Both metrics use complete
sentences; aggregate scores pool edit counts and reference lengths.

Across these cases, CER decreases from 21.21\% to 1.52\%, and ASR-derived
PSER decreases from 13.64\% to 0.00\%. The remaining character substitution
in ZH-02, ``赖'' versus ``癞'', maps to the same syllable, \texttt{lai4},
and therefore contributes no PSER error. These results indicate agreement
with the reference at the level of converted Pinyin, rather than direct
phonetic verification of the generated audio.

The interventions alter how the same character sequence is represented.
For example, ``贡献力量'' changes from one token to two lexical tokens
separated by a comma, while ZH-02 changes from ``武汉 / 麻 / 将以'' to
``武汉 / 麻将 / ， / 以''.
This suggests that the organization of text conditioning can affect
pronunciation. However, commas also change contextual representations and
may affect phrasing, so these cases do not isolate tokenization as the
cause. The edits are diagnostic, with no claim of improved prosodic
naturalness or generalization beyond the selected examples.
Our exploratory English analysis also found higher ASR word-error rates
among words represented by multiple BPE tokens. This association does not
separate tokenization from word length, frequency, or proper-name status.
\end{CJK*}

\end{document}